\documentclass{amsproc}

\usepackage{amsmath}
\usepackage{amssymb}
\usepackage{amsthm}
\usepackage{graphicx}
\usepackage[hidelinks=true]{hyperref}
\usepackage{cleveref}
\usepackage{xcolor}
\usepackage{mathtools}
\usepackage{todonotes}
\usepackage{fancyhdr}

\newtheorem{thm}{Theorem}

\theoremstyle{remark}

\definecolor{todo-color}{RGB}{120,205,205}
\newcounter{todo}

\numberwithin{equation}{section}

\begin{document}

\title[A tutorial on networks in social systems]{A tutorial on networks in social systems: A mathematical modeling perspective} 

\author{Heather Z. Brooks}
\address{Department of Mathematics, Harvey Mudd College, Claremont, CA 91711}
\email{hzinnbrooks@g.hmc.edu}
\thanks{The author gratefully acknowledges her co-organizers Michelle Feng, Alexandria Volkening, and Mason Porter, along with the American Mathematical Society, for their support and dedication in organizing this short course. This article was also informed and improved by the many thoughtful comments and questions from the participants of the 2021 short course. The author was supported in part by NSF Grant \#2109239.}

%    General info
\date{\today}


\begin{abstract}
This article serves as an introduction to the study of networks of social systems. First, we introduce the reader to key mathematical tools to study social networks, including mathematical representations of networks and essential terminology. We describe several network properties of interest and techniques for measuring these properties. We also discuss some popular generative models of networks and see how the study of these models provides insight into the mechanisms for the emergence of structural patterns. Throughout, we will highlight the patterns that commonly emerge in social networks. The goal is to provide an accessible, broad, and solid foundation for a reader who is new to the field so that they may confidently engage more deeply with the mathematical study of social networks. 
\end{abstract}

\maketitle

The study of complex systems is becoming an increasingly important area of inquiry \cite{bak2021stewardship}. With problems in epidemiology, misinformation on social media, voting and decision making, gerrymandering, and social justice appearing front-and-center in recent years, understanding the connections among people and among social entities is perhaps more important than ever. 

The study of networks is the study of connectivity. Networks encode the relationships among entities, and through this abstraction, we can study the structure and implications of these relationships. In this article, we will focus our attention specifically on networks that describe social systems. Such networks may describe social interactions and relationships that occur in physical spaces, such as the well-studied Zachary Karate Club network \cite{zachary1977information}. Networks may also be used to describe virtual connections as well, for example, Facebook friendships \cite{traud2012social}. It is worth noting that we need not limit our study to human social networks, as we observe a variety of interesting networks among other social animals as well (for example, the relationships of penguins at the Kyoto Aquarium \cite{Penguins_of_Kyoto_dataset}, grooming networks \cite{williams2018disease}, social spiders \cite{fisher2021using}, and more). At the time of writing, the Colorado Index of Complex Networks \cite{ICON} contains over 2000 network data sets on social systems.

The study of networks is an inherently interdisciplinary field, drawing from physics, biology, computer science, sociology, economics, and beyond. The tools of mathematics and mathematical modeling have an important role to play. In particular, by studying networks through a mathematical lens, we are well-positioned to explore some of the following big-picture questions in social networks:
\begin{enumerate}
	\item What network structures are likely to emerge in social systems? How can we interpret these structures?
	\item Why do particular network structures emerge?
	\item How do properties of networks affect the behavior and dynamics of social systems?
\end{enumerate} 

In this paper, my goal is to introduce the reader to some key mathematical tools to study social networks. In particular, we will focus on two areas: mathematical techniques for measuring network properties and generative models of networks. Along the way, we will highlight patterns that emerge in real social networks and provide you with resources that you can use to guide future study.

\section{Mathematical representations of networks}

We must first establish a foundation for how to describe and represent social networks mathematically. In this section, we discuss how graphs may be used to represent social networks, introduce some standard matrix representations of graphs, and describe some key terminology and properties. There are many excellent books on networks that describe each of the following properties in more detail. See, for example, \cite{newman2018networks, bullo2019lectures, jackson2010social}. These properties will also likely be familiar to those who have encountered graph theory. \footnote{The mathematical study of networks and the study of graph theory are, of course, deeply related. Both fields center around the graph as their primary mathematical object of study, and thus there are many overlaps and shared tools and methods. The difference between these subfields lies primarily in the questions of interest and the motivations behind those questions. A few introductory texts on graph theory include \cite{gould2012graph, gross2018graph, west2001introduction}} 

\subsection{Graphs}
We can represent networks with a graph $G = (V,E),$ where $V$ is the set of {\em vertices} or {\em nodes}, where the number of vertices in the graph is $\vert V \vert = n$. $E \subset V \times V$ is the set of {\em edges} between vertices. Vertices are represented visually with circles, and edges are represented as lines connecting those circles. The interpretation of vertices and edges in a graph to a particular social network setting is an important consideration from a mathematical modeling perspective. Often, vertices are chosen to represent the individuals or entities within a social network, and the edges are chosen to encode interactions, relationships, or other relevant connections between those individuals or entities.

\begin{figure}
	\includegraphics[width=1\textwidth]{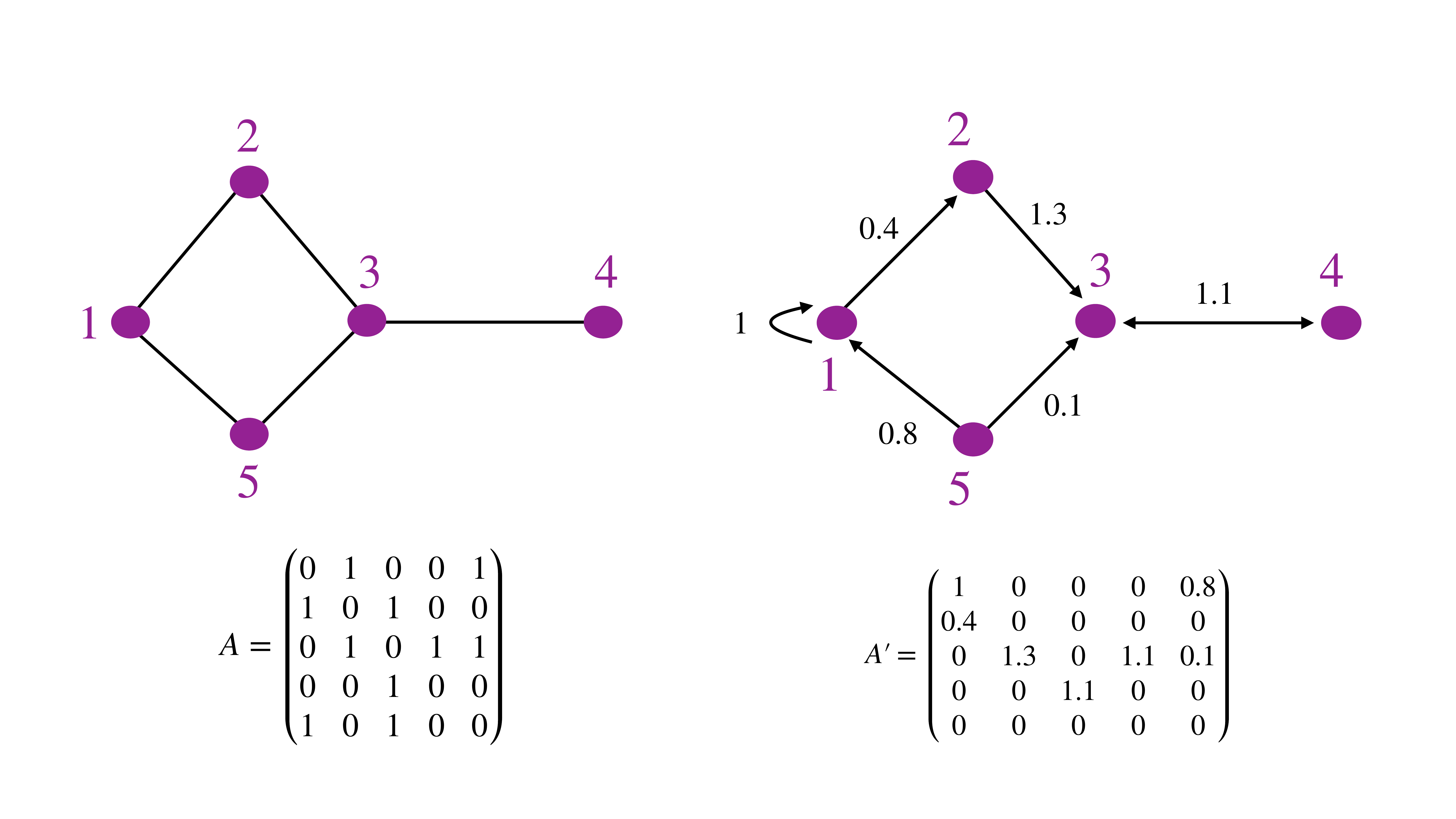}
	\caption{Two examples of graphs with their corresponding adjacency matrices. The graph on the left is a simple, undirected graph with 5 nodes and 5 edges. The matrix $A$ gives its adjacency matrix. The graph on the right is a directed, weighted graph with a self-loop on node 1. This graph has adjacency matrix $A'$; notice that the edge weights in this graph are encoded within the adjacency matrix.}
	\label{fig:graph-examples}
\end{figure}

{\em Simple graphs} are graphs that contain at most one edge between any distinct pair of vertices and no {\em self-edges} or {\em self-loops}, that is, no edges between a vertex and itself (see the example on the left in \cref{fig:graph-examples}). There are many useful extensions of this framework for the study of social networks. First, in some contexts, it may be helpful to allow self-edges to represent self-interactions. In many social networks, interactions and relationships may not necessarily be reciprocal. For example, one can follow an account on Twitter without being followed in return. To model this scenario, we often use a {\em directed graph}, where each edge has a directionality that is represented visually with an arrow. More formally, in a directed graph, the existence of the edge $(1,2) \in E$ does not imply the existence of the edge $(2,1) \in E.$ A graph that is not directed is said to be {\em undirected}. It may also be useful to relax the condition that any distinct pair of vertices are connected by at most one edge. To this end, we may choose to model our network with a {\em multigraph}, where we allow multiple edges (sometimes called {\em multiedges}) between each pair of nodes. A further generalization of this idea is to allow for each edge in a graph to have a weight, strength, or value (often rational or real-valued); such a graph is known as a {\em weighted graph}. For example, weighted graphs may be used to encode the distance between two points in a transportation network. In this scenario, one may choose edge weights that are inversely proportional to the geographic distance between the two points. \Cref{fig:graph-examples} shows two examples of graphs: one undirected simple graph and one directed, weighted graph with self-loops.

\subsection{Matrix representations of graphs}

One reason that graphs are particularly useful mathematical representations of networks is that they can be encoded with matrices. In this section, we will briefly introduce a few of the most commonly used matrix representations of graphs: the adjacency matrix, the incidence matrix, and the graph Laplacian.

\subsubsection{Adjacency matrix}
A graph $G$ with $n$ vertices can be represented by an $n \times n$ matrix ${\bf A}$ called the {\em adjacency matrix}, whose components $A_{ij}$ are defined as follows:
\begin{equation*}
A_{ij} = \begin{cases} w_{ij} & \text{if } (i,j) \in E \\ 
0 & \text{otherwise}\,,
\end{cases}
\end{equation*}
where $w_{ij}$ is the weight of edge $(i,j).$ A simple graph has an adjacency matrix that contains only ones and zeros, with a one in the $(i,j)$th component indicating an edge between node $i$ and node $j$. If a graph is undirected, its adjacency matrix will be symmetric.

\subsubsection{Incidence matrix}
While the adjacency matrix encodes relationships between node pairs, the {\em incidence matrix} encodes incidence between nodes and edges. If a graph $G$ has $n$ nodes and $m$ edges, the incidence matrix will be an $n \times m$ matrix ${\bf B}$ with entries
\begin{equation*}
B_{ij} = \begin{cases} w_{ij} & \text{if edge } j \text{ enters node } i, \\ 
-w_{ij} & \text{if edge } j \text{ leaves node } i, \\ 
0 & \text{otherwise}\,,
\end{cases}
\end{equation*}
where $w_{ij}$ is again the weight of edge $(i,j).$ Note that certain authors may flip the sign conventions. The incidence matrix is related to the adjacency matrix via the relationship ${\bf A} = {\bf B} {\bf B}^T - 2{\bf I}_n,$ where ${\bf I}_n$ is the $n \times n$ identity matrix.

\subsubsection{Graph Laplacian}
\label{sec:laplacian}
\begin{figure}
    \includegraphics[width=0.5\textwidth]{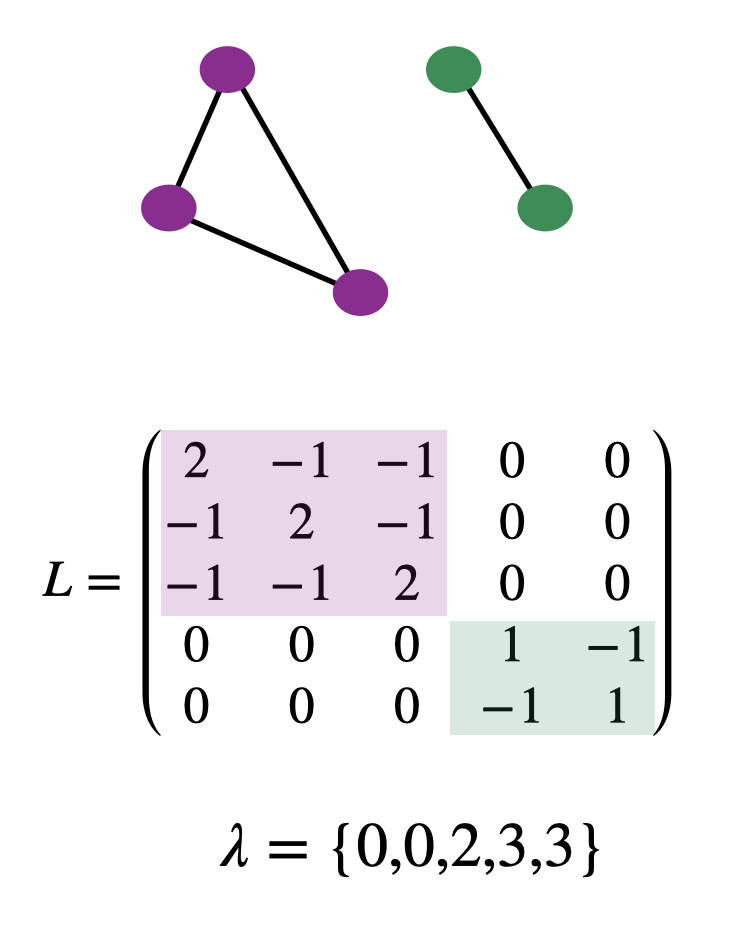}
    \caption{An example of a graph and its graph Laplacian $L$, along with the set of eigenvalues $\lambda$ of $L.$}
    \label{fig:laplacian}
\end{figure}

The graph Laplacian\footnote{There are many variants of the graph Laplacian. The version presented here is a commonly used unnormalized version. The way of defining the graph Laplacian --- including the choice of normalization --- may have impacts on applications, such as consistency in spectral clustering \cite{von2008consistency}.} is defined as 
\begin{equation}
\label{eqn:graphLaplacian}
    L = D - A \,,
\end{equation}
where $A$ is the adjacency matrix and $D$ is the diagonal matrix with entries $D_{ii} = \sum_{j=1}^n A_{ij}\,.$ See \Cref{fig:laplacian} for an example of the graph Laplacian for a graph with 5 nodes.
The graph Laplacian arises in many contexts, but it is particularly useful in the study of social networks due to its spectral properties. First, we note that all of the eigenvalues of $L$ are nonnegative, and the smallest eigenvalue is zero. The algebraic multiplicity of the zero eigenvalue allows us to infer the number of components of a network: a network has $c$ components if and only if $L$ has $c$ zero eigenvalues (see \Cref{sec:comp} and \Cref{sec:large-comp} for more information on components). The first nonzero eigenvalue is called the {\em algebraic connectivity} or {\em Fiedler eigenvalue} and can be used as a way to measure how ``well-connected'' the overall graph is. The associated eigenvector can be used to partition the network.
Von Luxburg \cite{von2007tutorial} provides a tutorial on the graph Laplacian and its applications to clustering problems, which includes proofs of the properties described above.  

\subsection{Terminology and basic properties}
With our mathematical representations of networks in hand, we may now begin to introduce some graph terminology that will help illuminate some basic properties of networks in social systems.

\subsubsection{Degree} \label{sec:degree}
One natural question that arises in the context of social networks is ``How many contacts (or interactions, or connections, etc.) does each individual in a network have?'' The answer to this question is quantified by the {\em degree} of a node, that is, is the number of neighbors adjacent to that node. More formally, in an undirected network, the degree $k_i$ of node $i$ is easily calculated from the adjacency matrix:
\[ k_i = \sum_{j=1}^n A_{ij} \,. \]
In a directed network, this measurement requires somewhat more subtlety. To distinguish between ingoing and outgoing edges of a node, we may define the corresponding notions of {\em in-degree} and {\em out-degree}. The in-degree of node $i$ is \[k_i^{in} = \sum_{j=1}^nA_{ij},\] that is, the number of ingoing edges of node $i$; the out-degree of node $i$ is \[k_i^{out} = \sum_{i=1}^nA_{ij},\] the number of outgoing edges of the same node. A familiar example of the use of degree, in-degree, and out-degree can be found in online social media platforms: if we model followerships on Twitter as a directed network so that an ingoing edge represents a follower, we can construe the number of accounts one is following as the out-degree and the number of followers of that account as the in-degree. By contrast, if we consider modeling Facebook friendships (which are reciprocal) as an undirected network, we may construe the degree of an account to represent the number of Facebook friends of a particular account.

\subsubsection{Paths, walks, and cycles} \label{sec:path}
\begin{figure}
    \includegraphics[width=0.5\textwidth]{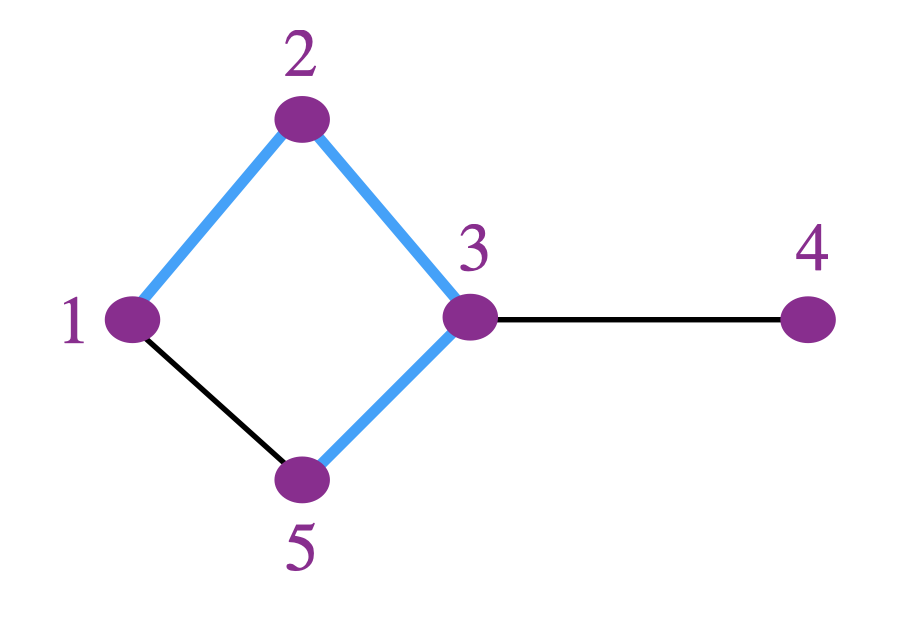}
    \caption{An example of a path from node 1 to node 5 is highlighted in blue. This path is of length 3 and consists of the sequence $\{ (1,2), (2,3), (3,5) \}$.} 
    \label{fig:path}
\end{figure}

A {\em walk} is a sequence of edges connecting a sequence of nodes. A {\em path} is a walk that does not intersect itself. \Cref{fig:path} shows an example of a walk.
Paths and walks can be used to study quantities like the geodesic distance between pairs of nodes; this will be discussed further in \cref{sec:small world}. The following theorem gives a convenient way to calculate the number of walks between two nodes using the adjacency matrix. This theorem is also useful for finding cycles (as cycles are walks from a node $i$ to itself).

\begin{thm}
    Suppose ${\bf A}$ is the adjacency matrix corresponding to a graph $G$ and $r$ is a positive integer. If ${\bf A}^r$ is the matrix product of $r$ copies of ${\bf A}$, then the number of walks of length $r$ between nodes $i$ and $j$ is the $(i,j)$th component of ${\bf A}^r.$
\end{thm}
To prove this theorem, proceed inductively on the walk length $r.$ Details of the proof may be found in many introductory texts on networks and graph theory.

\subsubsection{Components and connectivity}
\label{sec:comp}
\begin{figure}
    \includegraphics[width=\textwidth]{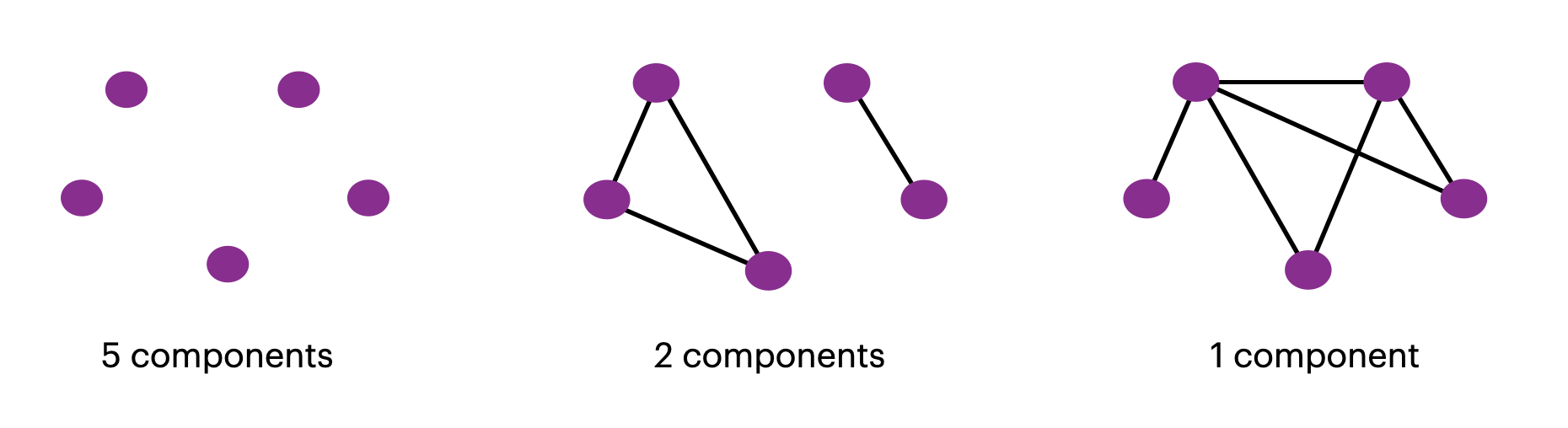}
    \caption{A graph with 5 nodes and 5 components (left), a graph with 5 nodes and 2 components (middle), and a graph with 5 nodes and one component (right). The graph on the right is connected.}
    \label{fig:comp}
\end{figure}

A {\em component} is a subset of nodes such that there exists at least one path between each pair of nodes in the subset such that no other node can be added to the subset and still preserve this property. A network in while all nodes belong to the same single component is {\em connected.} Some examples are given in \Cref{fig:comp}.

As discussed in \Cref{sec:laplacian}, the number of components in a graph corresponds to the number of zero eigenvalues in its graph Laplacian. This provides one straightforward method for finding components in a network.

\section{Properties of social networks}

\subsection{Social networks often contain a large component} 
\label{sec:large-comp}
In many social networks (and, in fact, in undirected networks in general), it is common for one component of the network to contain a large percentage of the nodes --- in many cases, upwards of 90\% of the nodes \cite{newman2018networks}. Why might social networks contain a large component? To understand which mechanisms might lead to this property, it is useful to study the emergence of large components in simple generative models of random graphs.

\subsubsection{The $G(n,p)$ model} \label{sec:Gnp}
Perhaps the simplest random graph model is the {\em $G(n,p)$ model} (sometimes also known as Erd\H{o}s--R\'enyi model). A realization of this model is created by fixing two parameters: $n$, the number of nodes, and $p$, the probability of an edge connecting any pair of nodes. Given $n$ and $p$, we may then generate a network by placing an edge between each distinct pair of nodes with independent probability $p$. Note that the number of edges  $m$ is not fixed --- using this strategy, we may generate a network with anywhere between $m=0$ and $m={n \choose 2}$ edges.  

\begin{figure}
    \includegraphics[width=1\textwidth]{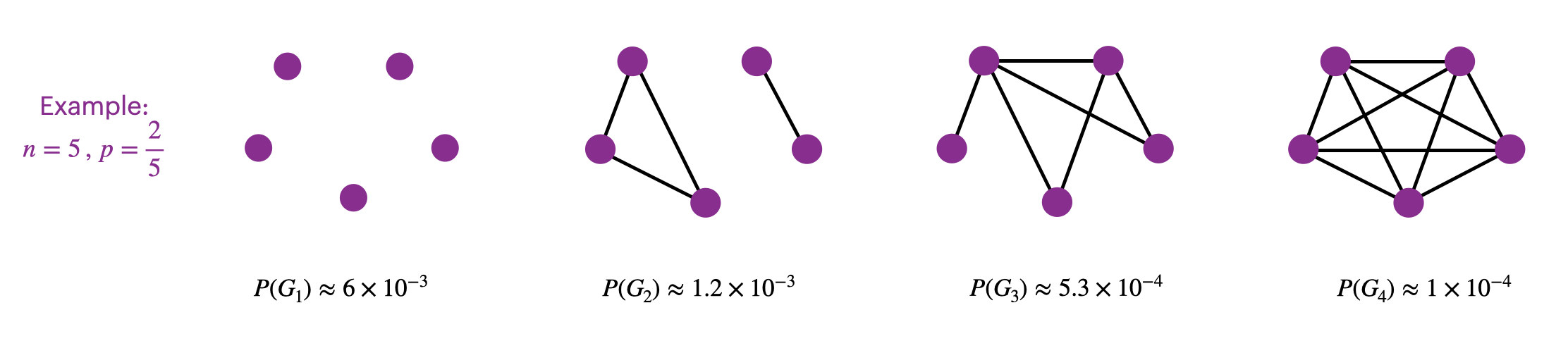}
    \caption{Calculating the probabilities $P(G_i)$ of a particular graph realization for a $G(n,p)$ model with $n=5$ nodes and an independent edge probability of $p=\frac{2}{5}.$}
    \label{fig:gnp}
\end{figure}

Mathematically, it is useful to think of a $G(n,p)$ model as an ensemble of simple networks with $n$ nodes, i.e., a probability distribution over possible graphs in which a graph $G$ appears with probability
\[ P(G) = p^m \left(1-p\right)^{{n \choose 2}-m} \,. \]
See \Cref{fig:gnp} for an example on a small network.

At this point, it is worth mentioning a related (but distinct) class of random graph models known as $G(n,m)$ model. In these models, we create a particular network realization by choose uniformly at random among the set of all simple graphs with $n$ nodes and $m$ edges. We will not explore $G(n,m)$ models further here; we will refer the interested reader to a networks textbook such as \cite{newman2018networks} for further details on the properties of these generative network models.

Let us return to the $G(n,p)$ model to develop some intuition behind the observed large components in social networks. Do networks generated with the $G(n,p)$ model contain a large component? If we ponder this for a few moments, we will realize that the answer to this question must be ``it depends.'' For example, if we consider $n=1000$ nodes and $p=0$, then the largest component of any network generated from this model will contain only one node (or 0.1\% of the nodes) --- certainly not an impressively-sized component! On the other hand, if $n=100$ and $p=1$, we will generate a network whose largest component contains all nodes with probability 1. Is there a transition between these two regimes?

To answer this question, rather than looking at the absolute size of the largest component, we will reframe the question as follows: for a given value of $p$, is there a component whose size grows in proportion to $n$? (If such a component exists, it is referred to as a {\em giant component}).  Following an argument from Erd\H{o}s and R\'enyi, we will show that for the $G(n,p)$ model there is a critical value of $p$ beyond which we expect the emergence of a giant component.

Suppose we let $u$ represent the average fraction of nodes that do not belong to this giant component. We know that, if node $i$ is not in the giant component, then it must not be connected to any node in the giant component, i.e., every other node $j \neq i$ is either not connected to node $i$ (with probability $1-p$), or it is connected to $i$ and also not in the giant component (with probability $pu$). Combining these two observations, we can note that the probability of a node not being connected to the giant component via a particular node $j$ is $1-p + pu.$ Thus, the probability $u$ of a node not being connected to the giant component by any of the $n-1$ other nodes is
\[ u = \left(1-p+pu \right)^{n-1}.\]
We can rewrite the expression above by noting that the probability that any two nodes are adjacent is $p = \frac{ \langle k \rangle}{n-1},$ so
\begin{align}
 u = \left(1- \frac{ \langle k \rangle}{n-1}(1-u)\right)^{n-1} \,, \\
 \Rightarrow \ln u = (n-1) \ln \left(1- \frac{ \langle k \rangle}{n-1}(1-u)\right)\,.
\end{align}

Rewriting the expression in this way allows us to derive an approximate expression for $u$ by performing a Taylor expansion about $u = 0$ and neglecting higher-order terms. Doing so yields
\begin{align}
	\ln u \approx - (n-1) \frac{ \langle k \rangle}{n-1}(1-u) \,, \\
	\ln u \approx- \langle k \rangle \left(1-u\right) \,, \\
	\Rightarrow u \approx e^{-\langle k \rangle (1-u)} \,.
\end{align}
Denoting the average fraction of nodes in the giant component by $S = 1-u,$ we obtain the following equation:
\begin{equation}
\label{eqn:S}
	S = 1- e^{-\langle k \rangle S}\,.
\end{equation}
While \cref{eqn:S} does not admit a simple closed form solution, we can use graphical methods to find solutions by looking for intersections of $y=S$ and $y= 1- e^{-\langle k \rangle S}.$

\begin{figure}
    \includegraphics[width = 0.8\textwidth]{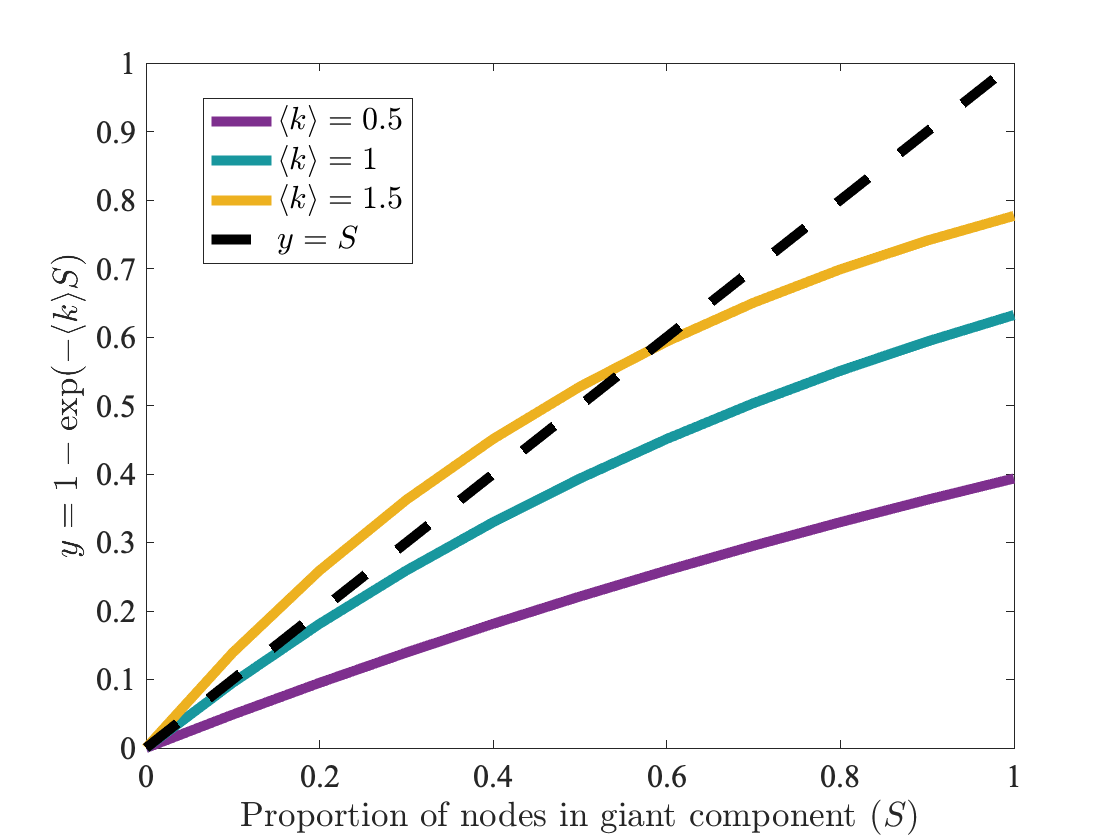}
    \caption{The expected proportion of nodes in a giant component ($S$) for a $G(n,p)$ random graph model is represented graphically with the intersection of $y=S$ (dashed black line) and $y= 1- e^{-\langle k \rangle S}.$ When the mean degree $\langle k \rangle$ is small, the only intersection is at the origin and we do not expect the emergence of a giant component (purple curve). As the mean degree $\langle k \rangle$ becomes large, an additional intersection point appears, signaling the possibility of the emergence of a giant component (yellow curve). The transition point between these regimes occurs at $\langle k \rangle = 1$ (teal curve).}
    \label{fig:giant-comp}
\end{figure}

First, we see by inspection that $S=0$ will always be a solution, that is, it is always possible that a particular network has no giant component. We instead turn our attention to determining when a giant component is possible. When the mean degree $\langle k \rangle$ is small, $S=0$ is the only solution to \cref{eqn:S}, however, for large $\langle k \rangle$, there is an additional intersection point where $S>0.$ For what value of $\langle k \rangle$ does this transition occur? \Cref{fig:giant-comp} gives us the necessary intuition to answer this question: to obtain two intersection points, it must be the case that $1- e^{-\langle k \rangle S}$ is growing faster than $S$ at $S=0.$
In particular, the transition from one intersection to two occurs precisely when these two expressions have equal growth rate at $S=0.$ That is,
\begin{align}
	\frac{d}{ds} \left( 1- e^{-\langle k \rangle S} \right) \bigg \vert_{S=0} = 1 \,, \\
	\Rightarrow \langle k \rangle e^{-\langle k \rangle S} \bigg \vert_{S=0} = 1 \,, \\
	\Rightarrow \langle k \rangle = 1.
\end{align}
This means that there can be no giant component for $\langle k \rangle <1.$ This calculation confirms our intuition: if the mean degree of a network is less than one, there are (relatively) few edges and it is quite easy for nodes to be disconnected from one another; we might expect that we have many small connected components. 

It remains to show that the $S > 0$ solution when $\langle k \rangle > 1$ is observed in practice. Using straightforward probabilistic arguments, one can show that as $n$ grows large, we do in fact expect to see a giant component emerge, and $S \rightarrow 1$ as $\langle k \rangle$ grows (that is, the number of nodes in this component approaches the number of nodes in the network).
 
\subsection{Degree distributions of social networks are often heavy-tailed}

\subsubsection{Degree-based centrality measures and degree distributions} \label{sec:deg-centrality}
{\em Centrality measures} quantify the relative importance of nodes in a network. We have already seen one centrality measure in \cref{sec:degree}: if we suppose that nodes with more connections are more important or influential in our network, then indeed the degree of a node can be used as a centrality measure. Social media networks like Twitter provide a good example: we may suppose that an account is more influential than another if it has more followers. It is important to notice that centrality is a relative measure; a node's centrality score only has meaning in comparison with another node.  

If $k_i$ is the degree of node $i$, then the set $\{k_1, k_2, \dots, k_n \}$ is called the {\em degree sequence} of a network; if we order the set from largest to smallest then we can see that the node(s) $j$ with the largest $k_j$ has the highest centrality, and we can continue the comparison. This method of calculating a node's centrality has a clear advantage in that it is easy to interpret and calculate: if we let ${\bf c}_{deg}$ be the vector containing the centralities of each node, then ${\bf c}_{deg} = {\bf A 1},$ where ${\bf 1}$ is the vector of ones.

One might argue that this simple centrality measure misses a key feature of relative importance: perhaps a node's centrality should be based not only on the number of connections it has, but whether or not it is connected to other important nodes (colloquially, ``it's who you know''). To incorporate this idea, supposing that the centrality of node $i$ is proportional to the sum of the centralities of its neighbors yields
\begin{align}
	c_i = \frac{1}{r} \sum_{j=1}^n {\bf A}_{ij}c_j \,,
\end{align}
where $r$ is a proportionality constant. A quick manipulation of this equation leads us to see that ${\bf c}$ is an eigenvector of ${\bf A}$ with eigenvalue $r.$ For this reason, this centrality measure is called {\em eigenvector centrality}. Which eigenvector should we use? Provided that the graph encoded by this adjacency matrix is connected, then an application of the Perron--Frobenius theorem tells us that ${\bf A}$ has a unique largest eigenvalue $r$, and its associated eigenvector contains only positive components.

There are some complications with applying eigenvector centrality for directed graphs or graphs with multiple connected components. The latter could be addressed by calculating centralities for each component separately (thus guaranteeing that the conditions of the Perron--Frobenius theorem are still satisfied). The former requires some more subtle modeling choices: should this centrality measure use in-edges or out-edges to calculate centrality? This will depend on the context of one's problem. It is in the context of directed graphs that we encounter a downside of eigenvector centrality: it is not difficult to construct networks where every node has zero eigenvector centrality, because (for example) a node with in-degree zero ``propagates'' its zero centrality throughout the network. Needless to say, eigenvector centrality does not provide a very useful measure in such a situation.

A fix for this is provided by {\em Katz centrality}, where we modify the eigenvector centrality by adding a baseline amount $\beta > 0$ to the centrality of each node, that is, ${\bf c} = \alpha{\bf A}{\bf c} + \beta {\bf 1}$. We are able to solve for ${\bf c}$ provided that ${\bf I} - \alpha {\bf A}$ is invertible, that is, provided that we choose $0< \alpha < \frac{1}{r},$ where again $r$ is the largest eigenvalue of the adjacency matrix ${\bf A}$. 

One practical consideration when choosing to implement Katz centrality is that if a node with high Katz centrality is connected to many other nodes, all of those adjacent nodes will end up with a high centrality score as well, which may be undesirable in certain contexts. One possible fix for this would be to ``dilute'' the centrality based on the number of out-edges a node has, that is, the centrality contributed by node $i$ is divided evenly among all of its adjacent nodes $j$. This modification results in the centrality equation
\begin{align}
	{\bf c} = \alpha {\bf A} {\bf D}^{-1}{\bf c} + \beta {\bf 1} \,,
\end{align}
where $D$ is the diagonal matrix with elements $D_{ii} = \max(k_i^{out}, 1).$ This is called {\em PageRank centrality} due to its connection to Google's search algorithms. Again, provided that $\alpha$ is chosen so that $I-\alpha {\bf A} {\bf D}^{-1}$ is invertible, this will yield a unique centrality ranking for a network's nodes.  There is another appealing interpretation of PageRank: Noting that ${\bf A} {\bf D}^{-1}$ is row stochastic, we can view this matrix as the transition matrix of a random walker traversing edges in our network. The PageRank centrality ${\bf c}$ is the stationary distribution for this Markov Process. For more details and applications of PageRank, see \cite{bryan200625, gleich2015pagerank}.

We conclude this section by emphasizing that there is no ``best'' centrality measure: the choice of centrality measure is itself a modeling decision, and should be made with careful consideration of its relevance to the quantities that are desirable to measure for a particular application \cite{landherr2010critical}. Indeed, there are may centrality measures based on quantities other than degree: for example, in \cref{sec:path-centrality}, we will discuss another class of centrality measures based on paths. Before doing so, we will explore degree distributions in social networks. Similar observations may be extended to other degree-based centrality measures.

\subsubsection{Degree distributions and the $G(n,p)$ model} \label{sec:deg-dist-Gnp}
First, we define the {\em degree distribution} of a network to be the function $p: \mathbb{N} \rightarrow \mathbb{R},$ where $p(k)$ is the number of nodes with degree $k$ divided by the total number of nodes $n$. In order to understand what we might expect from a degree distribution, we turn again to the simple $G(n,p)$ random graph model introduced in \cref{sec:Gnp}. Since these graphs are constructed under the assumption that the probability of any pair of nodes being connected is $p$, then the probability of a node being connected to $k$ other nodes (out of the possible $n-1$ other nodes in the network) is binomially distributed with probability $p$: that is, the degree distribution satisfies $p_k = {n-1 \choose k}p^k (1-p)^{n-1-k}.$ Furthermore, if $n >> 1$ and $p << 1$, the degree distribution is approximately Poisson: $p_k \approx e^{-\langle k \rangle}\frac{\langle k \rangle^k}{k!}.$

\subsubsection{Degree distributions in social networks} \label{sec:deg-dist-soc-networks}

What are the degree distributions observed in real social networks? We will briefly present three case studies (representing a much broader trend) that suggest that degree distributions are often heavy-tailed. That is, in contrast to the degree distributions observed in the $G(n,p)$ model, it is common in real networks that most nodes have a relatively small degree, while a small number of nodes have very large degree.

\begin{figure}
	\includegraphics[width=0.5\textwidth]{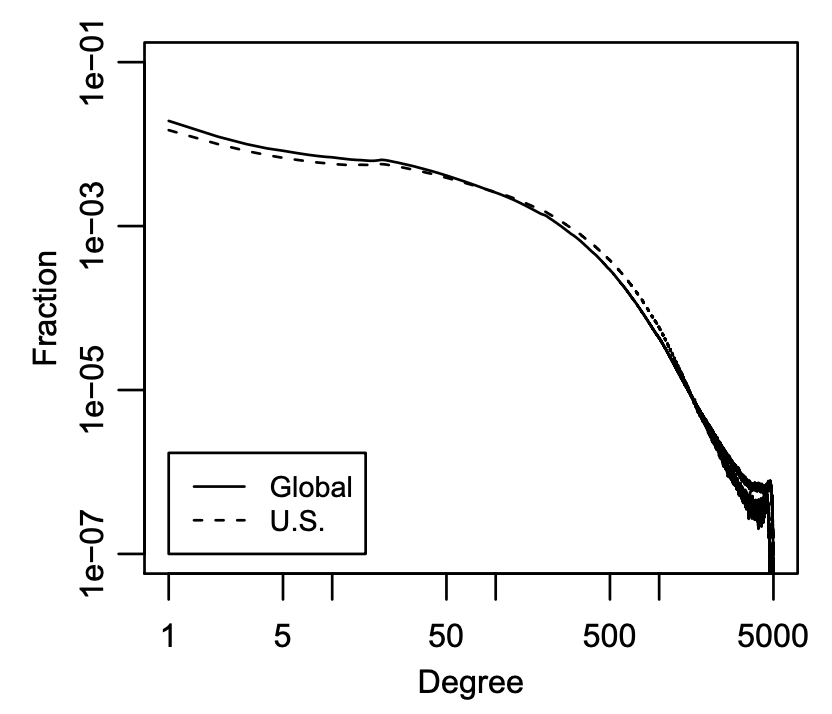}
	\caption{From \cite{ugander2011anatomy}: The degree distribution of Facebook users in 2011. Both globally and within the U.S., this degree distribution has a characteristic ``heavy tail'': Many users have relatively few friends, while a small proportion of users have a very large number of friends. Note that this distribution has a steep drop-off at 5000 due to an imposed limit on Facebook friends at the time of the study.}
	\label{fig:ugander-deg-dist}
\end{figure}

In Ugander et al.~\cite{ugander2011anatomy}, the authors study the structure of the network formed by users of Facebook, a popular social media platform. In this context, nodes represent users, with an edge occurring between any pair of nodes who are designated as friends on Facebook. This is a notable study in that it contains an impressive $n = 721$ million nodes and $68.7$ billion friendship edges. It is interesting to note that the authors report that 99.91\% of individuals belong to a single large connected component. The median number of Facebook friends for a user (i.e., the median degree) in this data set is $99.$ Looking at the degree distribution of this network gives a more detailed picture (\Cref{fig:ugander-deg-dist}): most individuals have a relatively small number of friends, while a small subset of users have thousands of friends. This is a characteristic example of a heavy-tailed degree distribution in a social network.

\begin{figure}
	\includegraphics[width=0.95\textwidth]{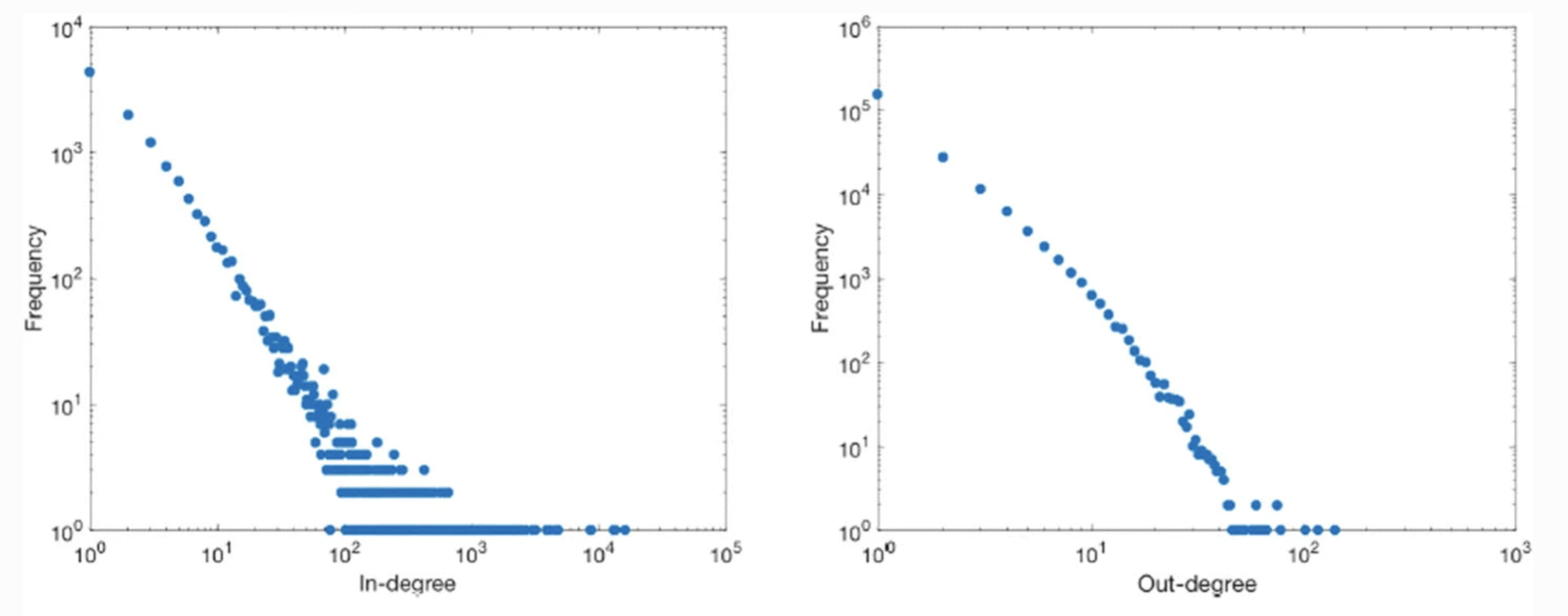}
	\caption{From \cite{tien2020online}: The in-degree distribution (left) and out-degree distribution (right) of a retweet network of Twitter messages with the hashtag \#Charlottesville. Both degree distributions have a characteristic ``heavy tail'': that is, a small number of accounts are retweeted much more often than most tweets in the dataset, and similarly a small number of accounts are responsible for a large proportion of retweets. While both distributions share this qualitative feature, we note that the tail of the in-degree distribution is much longer.}
	\label{fig:tien-deg-dist}
\end{figure}

We see heavy-tailed degree distributions continuing to show up in online social media networks today --- and not only in the scenario where edges are construed to represent friendships or followerships. To consider a more modern example, in \cite{tien2020online}, the authors study the `retweet network' of a dataset of messages (`tweets') from the social media network Twitter with the hashtag \#Charlottesville. They create a retweet network by representing accounts as nodes and using weighted, directed edges to represent the number of times account $j$ retweeted account $i$ (that is, account $j$ shared a message originally posted by account $i$). Note that accounts do not need to have any followership relationship to be able to retweet. As this network is directed, the authors examine the degree distributions for both in-degree (number of times an account was retweeted) and out-degree (number of times a node posted a retweet). Both distributions again have the characteristic heavy tail seen in other social networks (\Cref{fig:tien-deg-dist}).

\begin{figure}
	\includegraphics[width=0.95\textwidth]{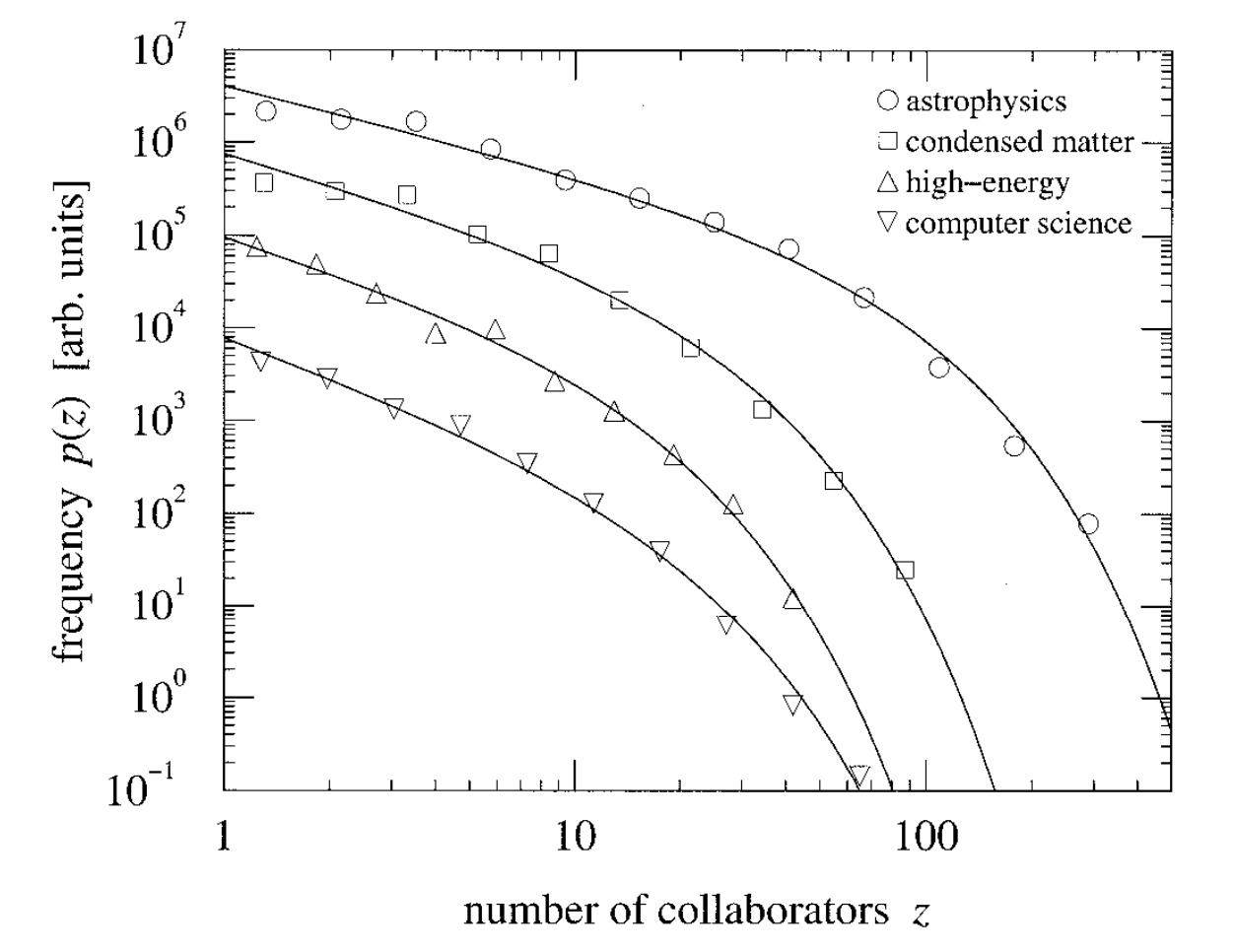}
	\caption{From \cite{newman2001structure}: Histograms of the number of collaborators (i.e., the degree in the co-authorship network) for authors of articles in astrophysics, condensed matter physics, high-energy physics, and computer science. The solid lines give a least-squares fit. All of these collaboration networks have a similar heavy-tailed degree distribution.}
	\label{fig:newman-deg-dist}
\end{figure}

Heavy-tailed degree distributions are prevalent in social networks outside of online social media platforms as well. A common object of study is the `co-authorship' or `collaboration' network in particular fields, where the nodes represent scientists or authors, and the edges represent whether two individuals have co-authored an article \cite{albert2002statistical}. In \cite{newman2001structure, newman2001scientific}, Newman studies co-authorship networks in physics, biomedical research, and computer science, finding a heavy-tailed degree distribution in each case (see \cref{fig:newman-deg-dist}). In \cite{barabasi2002evolution}, the authors study co-authorship networks of mathematicians and neuroscientists and observe similar results.

As we noted in \cref{sec:deg-dist-Gnp}, the $G(n,p)$ random graph model does not produce a network with a heavy-tailed distribution. How might we create a generative model that does reproduce this common feature of social networks? In 1967, Price introduced a so-called cumulative advantage model \cite{price1967networks, price1976general} for citations of journal articles, where new citations are accrued at a rate which is proportional to the current number of citations. Under these conditions, the distribution of the number of citations in the model is consistent with observations in real-world citations, namely, it follows an {\em inverse power law} or {\em Zipf law} (that is, probability of a node having $k$ citations satisfies $P(k) \sim k^{-\alpha}$ for $\alpha>0$. In this context, Price reported $\alpha$ in the range of 2.5--3). In 1999, Barabas\'{i} and Albert adapted this idea to the context of undirected networks to create their well-known model of {\em preferential attachment} \cite{barabasi1999emergence}.  This generative model can be constructed as follows. Starting with a small number of nodes $m_0$, add a new node at each time step with $m \leq m_0$ edge stubs. Each of those $m$ edge stubs should then be attached to an existing node $i$ in the network with probability $p_{k_i}$, where $k_i$ is the degree of node $i$. If $p_{k_i}$ is proportional to $k_i$, then we are in the context of preferential attachment: nodes with higher degree are more likely to gain new edges (sometimes colloquially referred to as a `rich get richer' scenario). After this process is repeated for many steps and a graph is grown via this preferential attachment mechanism, Barabas\'{i} and Albert showed that the resulting degree distributions again satisfy a power law, a common variant of a heavy-tailed distribution. Networks who degree distributions satisfy power laws are sometimes called {\em scale-free}. There has been vigorous debate in recent years over whether the real social networks with heavy-tailed degree distributions are in fact scale-free. While we will not give further details here, curious readers are encouraged to see \cite{holme2019rare} and references therein.

When seeking generative network models to create random networks that satisfy particular degree distributions, we need not limit ourselves to the preferential attachment or $G(n,p)$ models we have thus far described. {\em Configuration models} are a family of models of random graphs where the degree sequence or degree distribution is fixed (while these models are widespread in the networks literature today, see \cite{bollobas1980probabilistic} for an early theoretical study). One way to interpret a configuration model is that it's an ensemble of edge-stub matchings, where each matching with a given degree sequence occurs with equal probability (and any other matching outside of the given degree sequence occurs with probability zero). Alternatively, one can think about these models by fixing the degree distribution $p(k)$, and then a particular degree sequence $\{k_i\}$ occurs with probability $\prod_i p(k_i).$ Let us explore this family of models with a concrete example: the ensemble of random graphs with $\{k_i\}$ drawn from a Poisson distribution with mean $\lambda$ nearly recovers a $G(n,p)$ model for large $n$. However, these are not identical, as it is important to note that a configuration model that is generated through edge-stub matching may contain self-edges and multi-edges which do not occur in the standard $G(n,p)$ model.

\subsection{The ``small world'' effect: The diameter and/or mean geodesic distance of networks is often (surprisingly) small} \label{sec:small world}
In \cref{sec:path}, we introduced the notion of paths in graphs. In this section, we will explore the notable properties related to paths in social networks. One natural modeling question to ask is how closely connected any two individuals are within a network. While there are possibly many paths between two nodes in the same connected component, perhaps the most natural way to measure how closely connected two nodes are is to identify the {\em shortest path} (also known as the {\em geodesic distance}) between them. 
\begin{figure}
    \includegraphics[width=0.7\textwidth]{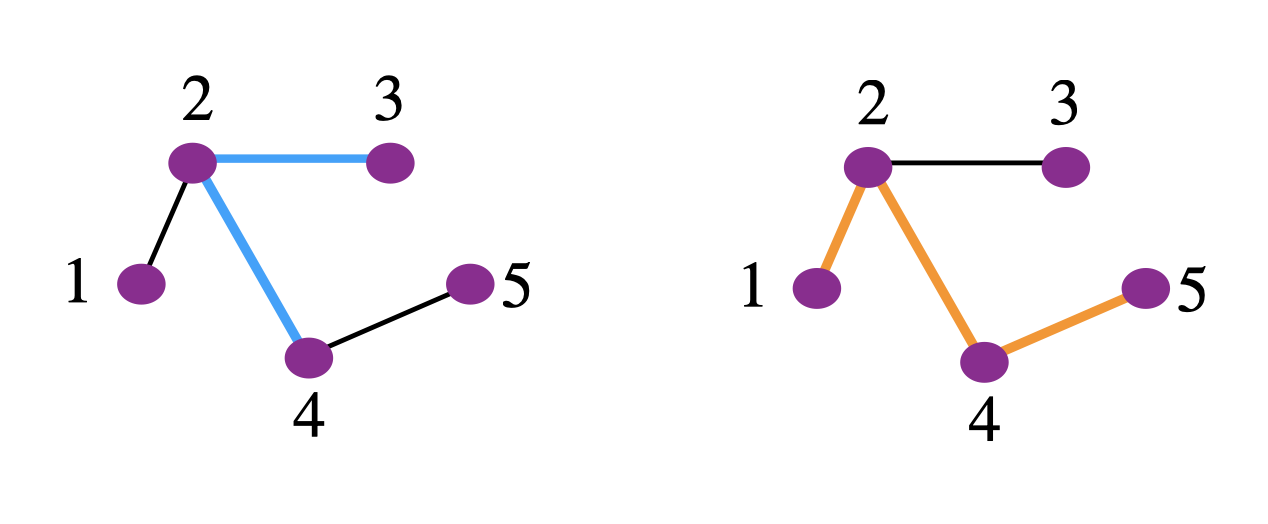}
    \caption{An schematic of shortest path and diameter in a network. The length of the shortest path between nodes 3 and 4 is $d_{34}=2$ (highlighted in blue, left). The diameter of the network is the longest shortest path in the network: this network has a diameter of 3, as $d_{15} = 3$ (highlighted in orange, right) and $d_{35}=3$.}
    \label{fig:path-diameter}
\end{figure}

When seeking to understand the topology of a full network, the notion of shortest paths gives some potentially insightful measures. One popular choice is to consider the mean shortest path length (sometimes called the {\em characteristic shortest path length}), which gives an idea of how many steps it takes on average to connect two nodes in a network. If we are instead more interested in the characterizing the extreme behavior (or `worst case scenarios'), we can use the {\em diameter} of the network, which is the longest shortest path in our network. Schematics of shortest path and diameter in a small network are shown in \Cref{fig:path-diameter}.

The concept of shortest path or diameter may already be familiar to you. In mathematics, we have a playful quantity called the Erd\H os number: a scholar who co-authored a paper with Erd\H os has an Erd\H os number of 1, a scholar who co-authored a paper with somebody who co-authored a paper with Erd\H os has an Erd\H os number of 2, and so on. In fact, we can see that the Erd\H os number is simply the shortest path in a co-authorship network between Erd\H os and another author in the same connected component. Of course, we do not need to limit our collaboration distance to Erd\H os. MathSciNet has a tool to compute the ``collaboration distance'' (i.e., the shortest path in the co-authorship network) between any two authors in their database \cite{mathscinet-collab}. Other (less math-centric) versions of this game include Six Degrees of Kevin Bacon (where the goal is find the shortest path between the actor Kevin Bacon to other actors in the `co-star' network, where edges occur between actors who have appeared in the same film) \cite{collins1998s} and Six Degrees of Wikipedia (where the goal is to find the shortest path between two Wikipedia pages through links) \cite{wiki-six-degs}. 

Each of the games mentioned in the previous paragraph are amusing examples of the {\em small world phenomenon}: the diameter (and/or mean geodesic distance) of social networks is often surprisingly small. This concept was popularized in Milgram's message-passing experiment, where participants were asked to try to deliver a letter to an arbitrarily selected person by sharing it with one of their personal acquaintances \cite{milgram1967small}. In revisiting our more modern example of the study of Facebook networks by Ugander et al.~\cite{ugander2011anatomy}, we see this play out in this online social network as well. The authors found that 99.6\% of users within their largest connected component had a `hop distance' (i.e., shortest path length) of six or fewer.

A quick back-of-the-envelope calculation can give some intuition as to why the small world phenomenon might occur in the 2011 Facebook friendship networks. Given that the median number of Facebook friends (median degree) is 99, we could estimate that the number of Facebook accounts within path length 6 of our own is approximately $99 \times 98 \times 98 \times 98 \times 98 \times 98 = 8.9 \times 10^{11},$ which exceeds the human population at that time (and even more comfortably exceeds the number of Facebook accounts).

While this calculation provides some helpful intuition, we must be quick to point out that we have neglected something very important: we have assumed that the set of our friends and the set of our friend's friends is distinct — surely we must have double-counted some individuals! We explore this feature further in the next section.

\subsubsection{Clustering and the clustering coefficient} \label{sec:clustering}
\begin{figure}
    \includegraphics[width=0.4\textwidth]{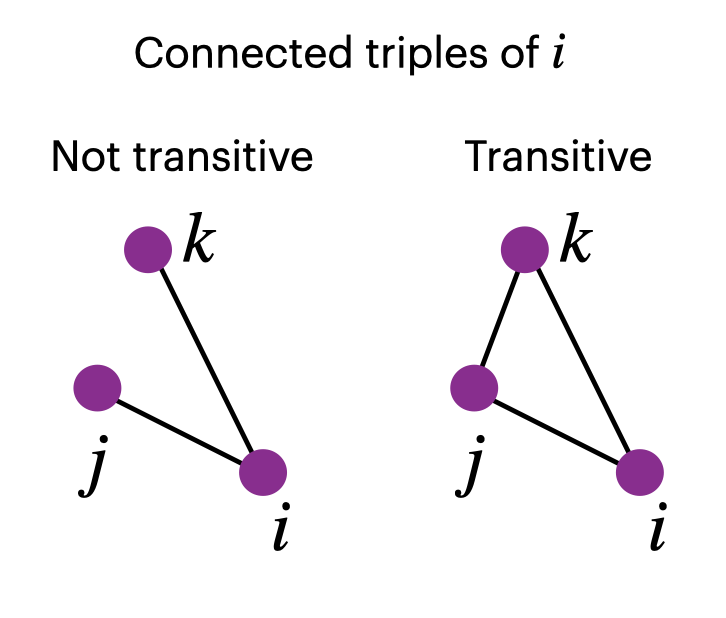}
    \caption{A schematic of connected triples of a node $i$. The example on the left is a connected triple, but is not a transitive triple because there is no edge between nodes $j$ and $k$. The example on the right is a transitive triple (or triangle), where edges exist between each node pairing of $i, j,$ and $k$. We can calculate the clustering coefficient by enumerating the transitive triples and the total number of connected triples and taking the ratio of these quantities.}
    \label{fig:triples}
\end{figure}

In our quick calculations from the last section, we neglected to account for whether or not two nodes that are adjacent to a node $i$ are also adjacent to each other. The extent to which this is true in a network is called {\em clustering} or {\em transitivity}. Colloquially, we might say in social networks that in a network with high clustering, ``the friend of my friend is also my friend.'' One way to measure local clustering for an individual node in a network is by defining the local clustering coefficient $C_i$, which is the ratio of the number of edges between neighbors of $i$ to the number of possible edges between neighbors of $i$. The {\em clustering coefficient} of a network can then be written $C = \langle C_i \rangle$, that is, the mean of the local clustering of all nodes in the network. Another alternative involves calculating the number of ``triangles'' or ``transitive triples'' in the graph, i.e., a trio of nodes that are all adjacent to each other (a complete subgraph with 3 nodes). The clustering coefficient for the network can then be defined by taking the ratio of the number of triangles in the graph to the total number of connected triples in the graph and multiplying by 3 (to account for the fact that each triangle gets counted three times as a connected triple). See \Cref{fig:triples} for a schematic of these quantities.

Real social networks can often exhibit high clustering. In our previous example with collaboration networks \cite{newman2001structure}, it was shown that two authors have at least a 30\% probability of collaborating with each other if they have both collaborated with a third author, and the values of clustering coefficient are higher than expected even when accounting for papers with 3 or more authors. \cite{ugander2011anatomy} find a relatively large clustering coefficient in their graph of Facebook friendships (0.14, i.e., on average 14\% of all the friend pairs of a median user are friends with each other). The authors of a large study \cite{myers2014information} of the Twitter followership graph containing 175 million users found a similar result to the Facebook study: nodes in the mutual graph (where an edge exists if two accounts follow each other) with degree 100 also have a clustering coefficient of approximately 0.14.

\subsubsection{Watts--Strogatz model} \label{sec:ws-model}
As we have explored in the previous two sections, real social networks are often characterized by both small characteristic path lengths and high clustering. Watts and Strogatz set out to create a generative random graph model that could capture both of these features \cite{watts1998collective}. Their idea is as follows: starting with a ring lattice of $n$ nodes with $k$ neighbors each, rewire each edge to a different end node with probability $p$ (disallowing duplicate re-wirings). This process effectively creates  ``shortcuts'' to decrease path length, while potentially retaining some of the clustering features from the original ring lattice. The parameter $p$ in this model can be tuned to balance these properties: when $p=0$, we produce a ring lattice (with high clustering, but relatively long path lengths), and when $p=1$, we recover a $G(n,m)$ model (with low clustering but a small mean shortest path). Watts and Strogatz show that as $p$ is increased from 0, the effect of the shortcuts ensures that mean shortest path decreases rapidly while the mean clustering coefficient remains high until the rewiring probabilities become larger. In particular, for values of $p$ on the order of 0.1, the resulting graphs tend to have relatively low mean shortest path length (roughly 20\% of the mean shortest path of the original ring lattice), but with clustering coefficient nearly as high as the original ring lattice. Their `small world' generative graph model, today often known as the Watts--Strogatz model, remains popular and well-studied.

\subsubsection{Path-based centrality measures} \label{sec:path-centrality}
We conclude this section on path-based properties of real social networks by returning to our earlier discussion about centrality (\cref{sec:deg-centrality}). Given what we now know about the features of shortest paths in social networks, we may be inspired to quantify the relative importance of a node in our network not by its number of connections, but by its path-related features. 

One possibility is to consider a node to have high centrality if it is a short `distance' from many other nodes, i.e., its mean shortest path length to other nodes is small. This is called {\em closeness centrality}, and one way to write this is to suppose that the centrality is inversely proportional to the mean shortest path to other nodes: If $d_{ij}$ is the shortest path from node $i$ to node $j$, then the closeness of node $i$ is $c_i = \frac{n}{\sum_{j \neq i} d_{ij}}.$ We need to take careful consideration when defining closeness in directed networks or networks with multiple connected components, however. This problem can be solved by considering closeness in each strongly connected components separately (although the scores will be not be comparable between components, as the size of the component affects these values). Another option is to redefine closeness in terms of the harmonic mean, so that $c_i = \frac{1}{n-1} \sum_{j \neq i} \frac{1}{d_{ij}}$, allowing that terms where there is no path between two nodes will contribute $0$ to a node's centrality (in some sense, this is like supposing $d_{ij}$ is infinite if there is no path between $i$ and $j$).

Another possibility is to suppose a node is instead important if it lies on many shortest paths between other nodes. For example, such nodes may be important because information or goods would pass through these nodes frequently, and their removal could disrupt paths. There are multiple ways to define such a centrality, which is known as {\em betweenness centrality}. One straightforward way to calculate this is to sum the number of all shortest paths of a pair of nodes that contain node $i$, and then normalize by the number of shortest paths between that pair. Summing over all possible pairs in the networks yields a betweenness centrality score. 

The centrality scores mentioned in this article are among the most common, but many possible variants abound in the literature. Indeed, you may find that your problem of interest requires the invention of yet a new strategy to define centrality. In selecting a centrality score, it is important to consider the benefits and pitfalls of a given method. Boldi and Vigna \cite{boldi2014axioms} provide a thorough discussion on centrality scores and suggest axioms for centrality measures.

\subsection{Assortativity, clustering, and community structure in social networks} \label{sec:assortativity}

\subsubsection{Assortativity and homophily}. The tendency of people to associate with others whom they perceive to be like themselves is called {\em homophily}. Supposing that we have a network where nodes can be categorized by different classes, types, or groups, we may like to develop a way to quantify the prevalence of homophily in our network. Informally speaking, a network is said to be {\em assortative} if a significant fraction of edges are between nodes of the same `type' and {\em disassortative} if a significant fraction of edges are between nodes of different `types.' For example, in a network of romantic relationships of high school students \cite{bearman2004chains}, the study's authors found this network to be disassortative by gender.

One way to quantify the assortativity of a network is to measure the {\em modularity} of the network. Suppose that we have a graph {\bf G} with a set of node classes or types $\mathcal{C}.$ Let $g_i \in \mathcal{C}$ be the type of node $i$, and $m$ be the number of edges in the network. Then the modularity $Q$ is defined to be 
\begin{equation}
\label{eqn:modularity1}
	Q = \frac{1}{2m} \sum_{i,j} \left(A_{ij} - \frac{k_ik_j}{2m} \right) \delta_{g_ig_j} \,,
\end{equation}
where $k_i$ is the degree of node $i$ and $\delta_{(\cdot, \cdot)}$ is the Kronecker delta. Let us quickly deconstruct the meaning of this expression. Notice that the first term will give us the fraction of same-type edges in our network. In the second term, we notice the quantity $\frac{k_ik_j}{2m}$, which will give approximately the expected number of edges between nodes $i$ and $j$. Thus, the second term gives the expected fraction of edges between all node pairs. In this way, \cref{eqn:modularity1} gives a comparison between the observed fraction of edges between same-type nodes and the expected fraction of edges between same-type nodes, given that they were connected randomly (i.e., via a configuration model as described in \cref{sec:deg-dist-soc-networks}). If $Q>0,$ then we observe more edges between same-type nodes than would be expected by chance; this corresponds to assortative mixing. If $Q<0$, then conversely the network is disassortative. 

There is an alternative definition of modularity that can provide easier calculation if we don't have information about the degree of individual nodes. First, we can define the fraction of edges that join nodes of type $g$:
\begin{equation*}
	e_g = \frac{1}{2m}\sum_{i,j} A_{i,j}\delta_{g_i,g}\delta_{g_j,g} \,.
\end{equation*}
The fraction of ``ends'' of edges attached to nodes (that is, node stubs) of type $g$ is
\begin{equation*}
	a_g = \frac{1}{2m} \sum_i k_i \delta_{g_i,g} \,.
\end{equation*}
Via some algebraic manipulation, we can rewrite $Q$ from \cref{eqn:modularity1} as
\begin{equation}
	Q = \sum_g \left(e_g - a_g^2\right) \,.
\end{equation}

Depending on the node types or classes under consideration, real-world social networks may be either assortative, disassortative, or neither. However, there is one important feature that is shared across many social networks: they tend to be assortative by degree (i.e., ``degree assortative'') \cite{newman2002assortative}. This means that in these networks, high-degree nodes are more frequently attached to other high-degree nodes than one would expect were edge stubs connected randomly. In \cite{newman2002assortative}, Newman examines a variety of real-world networks and observes degree assortativity in a variety of social networks, including co-authorship networks, film actor collaborations, and connections between business people. This feature is not true of all real-world networks: non-social networks such a protein interactions in yeast, synaptic connections in {\em C. elegans}, and food webs in various aquatic environments are disassortative by degree.

\subsubsection{Community structure} \label{sec:community}
\begin{figure}
    \includegraphics[width=0.8\textwidth]{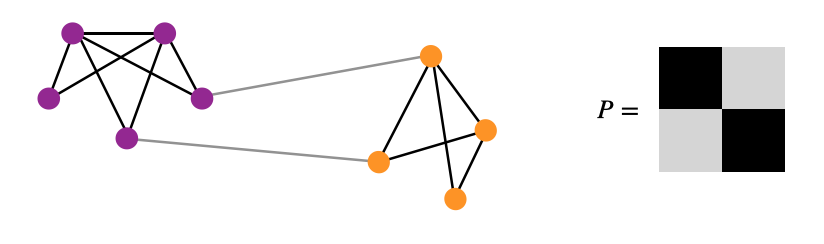}
    \caption{A schematic of a small network with community structure. The subgraphs with purple and orange nodes are densely connected (black edges), while there are fewer links connecting nodes between these purple/orange subgraphs (gray edges). Such an example can be generated by a stochastic block model with a matrix $P$ of connection probabilities like the one visualized above: the diagonal blocks $P_{ii}, P{jj}$ represent higher connection probability between nodes in the same community (black), and the off-diagonal blocks $P_{ij}, P_{ji}$ represent lower connection probability between nodes in different groups (lighter gray).}
    \label{fig:community}
\end{figure}

In \cref{sec:clustering} and \cref{sec:assortativity}, we have separately considered evidence suggesting that social networks exhibit high clustering and tend to be assortative by degree. In fact, these features of real social networks may arise from the same source: community structure. A graph with {\em community structure} consists of subgraphs (called ``communities'') which are more densely connected within those subgraphs than between them (see \Cref{fig:community}).
This somewhat simplistic definition can be made more precise in various ways. For example, one possibility is to define a community via the probabilities of edges existing between nodes within the community versus outside of it; this leads to the notions of {\em strong community} (each nodes within the subgraph has a higher probability to be connected to the nodes within the subgraph than to any outside of it) and {\em weak community} (each node's average probability is higher to be connected to other nodes within the subgraph than the average probability of the connection with nodes of any other group) \cite{fortunato2016community}. 

One of the most celebrated examples of the existence of community structure in social networks is the Zachary Karate Club network \cite{zachary1977information}. In his 1977 work, Zachary observed the social interactions of members of a karate club over the course of three years. During this time, members of this club became divided into two factions over a club-related issue. Zachary's insight was that, by characterizing the social interactions of the club members prior to this fission as a network, each faction corresponded to a more densely-connected subgroup of the network — thus, these two social `communities' (in our modern networks terminology) were in some sense predictive of which side individuals would take on the issue. This network has today become a common benchmark for algorithms for community detection, likely due to the presence of a ``ground truth'' (i.e., which faction the nodes would ultimately join) and its small size (34 nodes).

More recent examples of social networks with community structure abound. Examples include scientific collaboration networks \cite{newman2004finding}, where individuals in similar subfields are more densely connected; Facebook friendships of Caltech students, where connections were more dense between students between students from the same dormitory (`House') \cite{traud2012social}; and in the retweet network of individuals using a particular hashtag on Twitter; where accounts with similar media-followership behaviors retweet each other much more often \cite{tien2020online}.

Community detection in networks has been and remains a popular area of study, and methods abound to tackle this problem. Many of the topics we have addressed so far in this article form the foundations of these methods: some authors have used centrality-based methods (e.g., using betweenness as in the Girvan--Newman method \cite{girvan2002community}) and modularity optimization (e.g., with spectral partitioning via the graph Laplacian as in \cite{newman2006finding}). To survey community detection methods, benchmarks for community detection, and further generalizations, we refer the reader to \cite{porter2009communities, fortunato2016community}.

At this point, we meet our final class of generative models for this article. {\em Stochastic block models} are a family of random graph models that capture community structure. Suppose, as before, that we presume each node belongs to a particular group or type; the $i$th group is denoted $g_i.$ We can then generalize the ideas from the $G(n,p)$ model discussion in \cref{sec:Gnp} by defining the probability of an edge between to nodes depending on their respective types. That is, the probability of an edge between node $i$ and $j$ is $p_{g_ig_j}.$ If we define a matrix of connection probabilities $P_{ij} = p_{g_ig_j}$, this creates a block structure in this matrix $P$ (giving this model its name). This versatile model allows us to create random graphs with a variety of interesting features. First, note that the classical $G(n,p)$ model is a special case of this stochastic block model. We can also create a network with community structure (by defining e.g. $P_{ii} > P_{ij}$ and $P_{jj} > P_{ij}$) or disassortative networks (allowing $P_{ij}>P_{ii}$ and $P_{ij} > P_{jj}$). We can even create networks with {\em core-periphery structure}, where certain nodes (the {\em core}) have high degree relative to the remaining nodes (the {\em periphery}). Supposing that group $g_i$ denotes the core nodes, this is achieved by allowing $P_{ii} > P_{ij} > P_{jj}.$ For a visualization of a stochastic block model for generating community structure, see \Cref{fig:community}.

\section{Overview and conclusions}

In this article, we have explored several key features of the structure of social networks. We briefly review these properties below.
\begin{description}
 \item[A large component]  It is common in social networks for a high proportion of nodes to be in the same connected component (\Cref{sec:large-comp}).

 \item[Heavy-tailed degree distribution] Social networks often contain a small number of nodes whose degree is relatively high, and a larger number of nodes with small degree (\Cref{sec:deg-dist-soc-networks}).

 \item[Small diameter] In many social networks, the shortest path between any two nodes can be quite small; this is sometimes known as the {\em small-world property} (\Cref{sec:small world}).

 \item[High level of clustering] If two nodes $i$ and $j$ in a social network are each connected to the same third node $k$, it is more likely that $i$ and $j$ are also connected to each other (\Cref{sec:clustering}). 

 \item[Assortative by degree] Social networks may be assortative, disassortative, or neither, depending on the node classes or types under consideration. However, one frequently-shared property of social networks is that nodes with high degree are more likely to be connected to other nodes of high degree (\Cref{sec:assortativity}).

 \item[Community structure] Many social networks contain relatively densely-connected subgraphs called communities, with sparser connections between these communities (\Cref{sec:community}).
\end{description}

While it is interesting to measure and observe these properties empirically in real-world networks, analyzing generative mathematical models of networks can give insight into the possible mechanisms underlying the commonly-observed structural features listed above. In this article, we described several well-known generative models for networks to help illuminate these properties. 
\begin{description}
\item[$G(n,p)$ and $G(n,m)$ models] In $G(n,p)$ random graph models, the number of nodes is fixed, with each edge between a pair of nodes occuring with probability $p$. These models are simple to generate and are amenable to analysis (\Cref{sec:Gnp}).

\item[Preferential attachment models] Generative models of preferential attachment such as those by Price and Barabas\'{i}--Albert give us a way to create a network where the degree distribution of the nodes follows a power law as the number of nodes gets large. This is achieved by attaching new nodes to an existing node with a probability that is proportional to the degree of the existing node (\Cref{sec:deg-dist-soc-networks}).

\item[Configuration models] Configuration models are a family of models where the degree distribution or degree sequence is fixed and a network is generated that satisfies the given distribution. These models are popular choices due to their flexibility, their analytical convenience, and the ability to match degree distributions from real systems (\Cref{sec:deg-dist-soc-networks}).  

\item[Watts--Strogatz models] The Watts--Strogatz model is an example of a generative network model that can create networks with high levels of clustering and small shortest path lengths. This model is parameterized via a rewiring probability (\Cref{sec:ws-model}).

\item[Stochastic block models] Stochastic block models can generalize some of the previously described models to generate networks with community structure by allowing nodes to have heterogeneous connection probabilities between different groups or classes (\Cref{sec:community}).

\end{description}

The mathematical study of social networks is an exciting and growing field. There is much work to be done both in terms of understanding the applications and in terms of advancing the mathematics of networks \cite{porter2020nonlinearity+}. From a mathematical perspective, there is still much work to be done to understand the properties of time-dependent (or {\em temporal}) networks, that is, networks that change in time by adding, removing, or changing nodes and/or edges in time \cite{holme2012temporal, masuda2016guide}. Another active area of research is the theoretical study and application of dynamical systems on networks \cite{porter2016dynamical}. In such a system, each node has a state that changes in time (perhaps through interactions that are restricted via the network structure, e.g., occurring via edges in the network). There are many interesting properties to study in such systems: transient states, stationary states, stability analyses, bifurcations, and more. {\em Adaptive} or {\em co-evolving} network models combine these two ideas by studying networks where the node states and network structure interact and change in time \cite{sayama2013modeling, gross2006epidemic}. 

There are also several interesting structural generalizations worth noting. Multilayer networks \cite{kivela2014multilayer, porter2018multilayer, finn2021multilayer} provide a framework to represent different types of relationships or time-dependent relationships between nodes \cite{fisher2021using}. Research on higher-order networks that encode relationships beyond pairwise relationships includes the study of hypergraphs and simplicial complexes \cite{battiston2020networks, bianconi2021higher, benson2016higher, bick2021higher, mulas2022graphs}, which can be important for understanding structure and dynamics of social systems. The study of dynamical systems on these higher-order structures is still relatively young and there is much work to be done in these areas.

In this tutorial, we provided an overview of foundational topics in understanding structural properties and generative models of social networks. With these foundational techniques in hand, the hope is that the reader feels empowered continue to dive deeper into the exciting work in the mathematics of social networks.   

\bibliographystyle{amsalpha}
\bibliography{nssbib}

\end{document}